\documentclass[aps,twocolumn,superscriptaddress,showpacs]{revtex4}
\usepackage{graphicx}

\begin{document}

\title{Ultrathin optical switch based on a liquid crystal/silver nanoparticles mixture as a tunable indefinite medium}

\author{Elisa Spinozzi}
\address{University of Rome ``La Sapienza'', Department of Information Engineering Electronics and Telecommunications, Via Eudossiana 18, 00184 Roma,
Italy}

\author{Alessandro Ciattoni}
\address{Consiglio Nazionale delle Ricerche, CNR-SPIN, 67100 Coppito L'Aquila, Italy}
\email{alessandro.ciattoni@aquila.infn.it}

\begin{abstract}
We predict that a liquid crystal/silver nanoparticles mixture can be designed so that, in a frequency range, its effective ordinary and extraordinary
permittivities have real parts of different signs. We exploit this result to design a nano-photonic device obtained by sandwiching a few hundred
nanometer thick slab of the proposed mixture between two silica layers. By resorting to full-wave simulations, we show that, by varying the direction
of an externally applied electric field, the device can be used as an optical modulator since its transmissivity can be switched between $0.02$ and
$0.4$ at a wavelength close to the frequency range where the medium is indefinite. The device functionality physically stems from the fact the
orientation of the hyperbola characterizing extraordinary waves within the indefinite medium follows the applied electric field direction and
therefore, if the hyperbola asymptote is nearly normal to the slab, full switch between evanescent and homogeneous propagating waves can be achieved
within the medium.
\end{abstract}

\maketitle

\section{Introduction}
Within the rapidly growing field of metamaterials, indefinite media \cite{Smit1}, or extremely anisotropic metamaterials characterized by principal
permittivities having different signs, are attracting a considerable research interest mainly for the novel potentialities they provide for achieving
efficient optical steering and manipulation. The hyperbolic dispersion relation characterizing extraordinary plane waves through indefinite (or
hyperbolic) media is the main physical ingredient leading to unusual optical effects. Examples are negative refraction \cite{Smit2,Smit3,FangF}, due
to the fact that the hyperbola curvature is opposite to that of the circumference of standard media, and hyperlensing \cite{Jacob,LeeLe,YaoYa,Casse},
due to the possibility of converting vacuum evanescent waves into homogeneous propagating waves \cite{LiuL2} (as a consequence of the lack of
evanescent waves along the hyperbola major axis). Extreme anisotropy also fundamentally affects the excitation of waves, impinging from vacuum, into
an indefinite medium to the point that, by suitably changing the mutual geometric orientation between the hyperbola and the metamaterial interface,
novel interesting effects as cancellation of reflection and transmission can be predicted \cite{YangY}. Consequently, optical devices exploiting
indefinite media have been proposed as, for instance, polarization beam splitters \cite{Zhao1} and angular filters \cite{Aleks} which are based on
the fact that transverse magnetic and transverse electric waves are characterized by different dispersive features (hyperbolic and standard behavior,
respectively).

The condition of extreme anisotropy, where the principal permittivities have different signs, can be attained by resorting to suitable
metal-dielectric nano-composites with a geometrical structure exhibiting at least one privileged direction. Exploiting the effective medium theory,
it is possible to design the composite to show indefinite permittivity in a spectral range whose wavelengths are much greater than the sizes of the
nano-constituents. Two relevant examples of such structured media are nanowire composites \cite{Elser,LiuL1,Nogin} (typically silver nanowires
embedded in an allumina membrane) and layered media \cite{Schil,Hoffm,Huang,Korob} (obtained by alternating metal and dielectric layers) whose
privileged directions are, evidently, the nanowire and the stacking direction, respectively. Recently, it has been proposed that indefinite
permittivity can even be observed in anisotropic natural materials as, for example, graphite \cite{SunSu} and strong anisotropic uniaxial crystals
\cite{Wang1} which are shown to be indefinite in the ultraviolet (due to the hybrid electronic transitions) and the infrared region (due to the
strong anisotropy of polar lattice vibrations), respectively.

Dynamic light manipulation is a key ingredient of photonics and therefore investigating tunable metamaterials is an essential task. Basically, it has
been proposed that if the metamaterial constituents are immersed into a liquid crystal \cite{KhooK,Gorku,Prati} or the liquid crystal is used as one
of the metamaterial constituents \cite{Wang2,Werne,Minov}, the overall refractive index can be controlled through an external electric field, even
allowing the full switch of the refractive index sign. Although the present research interest in tunable metamaterials is rapidly growing, it should
be stressed that, to the best of our knowledge, an externally driven metamaterial with indefinite permittivity has not been considered yet.

\begin{figure*}
\centering
\includegraphics*[width=0.8\textwidth]{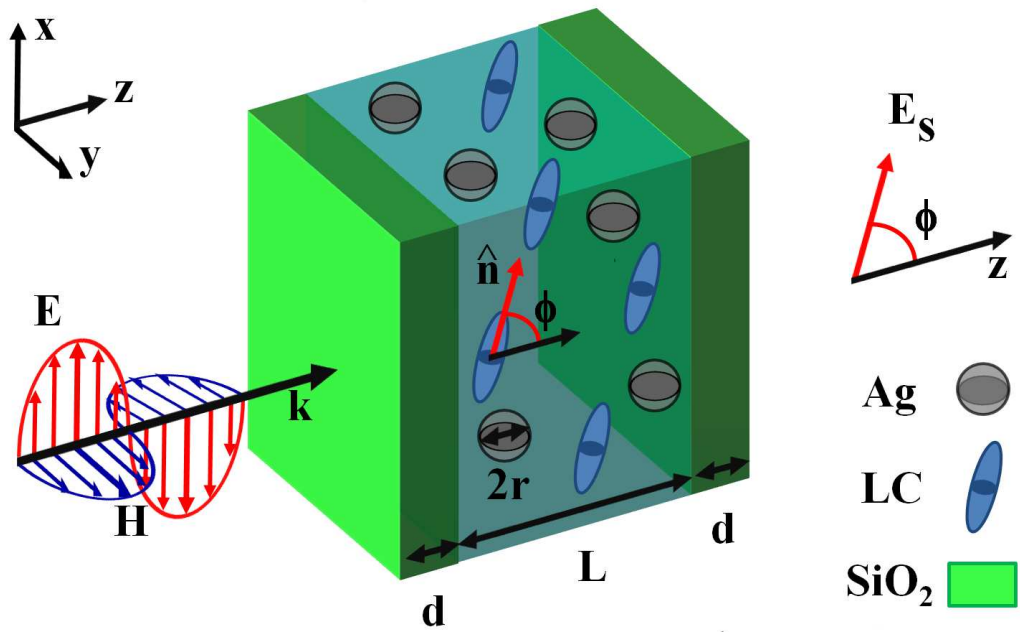}
\caption{Setup geometry. The liquid crystal (LC) fills the slab of thickness $L$, it is sandwiched between two silica layers of thickness $d$ and it
hosts silver nanoparticles (Ag) of radius $r$. The externally applied electric field ${\bf E}_S$, inclined at an angle $\phi$ with the $z$-axis,
aligns the liquid crystal molecules in the direction of the unit vector $\hat{\bf n}$ (parallel to the electric field). The vacuum electromagnetic
plane wave is made to impinge orthogonally onto the device interface.}
\end{figure*}

In this paper we consider a liquid crystal/silver nanoparticles mixture and we show that, by suitably choosing the volume filling fraction of the
particles, the anisotropic effective medium theory predicts the existence of a spectral range where the permittivity tensor is indefinite. Therefore,
to the best of our knowledge, we theoretically discuss the first example of indefinite optical metamaterial which 1) is obtained by dispersing metal
nanoparticles within an anisotropic matrix and 2) is tunable since, due to the presence of the liquid crystal, an external electric field is able to
change its optical response. These achievements allow us to consider an active optical device obtained by sandwiching a $300$ nm thick slab of the
considered tunable indefinite medium between two $70$ nm thick silica layers and operating in the presence of an externally applied electric field.
Exploiting the effective medium theory, we show that, at optical wavelengths where the inner medium is indefinite, the slab is completely opaque if
the electric field is orthogonal to the its interfaces whereas it becomes gradually more transparent if the angle is increased. Such a behavior is
physically due to the fact that the orientation of the hyperbola characterizing the dispersion relation of extraordinary waves through indefinite
medium follows the direction of the applied electric field and therefore, when the hyperbola asymptote is close to be normal to the device interface,
a switch between evanescent and propagating waves within the medium is achieved thus producing the change of the slab transmissivity. In order to
prove that the considered medium does actually exhibit indefinite permittivity and that the considered device operates the predicted optical switch,
we have performed full-wave (FDTD) simulations and evaluated the device transmissivity as a function of the wavelength and the applied electric field
direction. We show that a spectral region where the device operates the switch exists and it is very close to that predicted by the effective medium
theory and, most importantly, that, at a specific wavelength, the actual device transmissivity can be switched between $0.02$ and $0.4$ through the
externally applied electric field.

\section{Liquid Crystal/silver nanoparticles mixture as a tunable indefinite medium}
Consider a slab of thickness $L$ with its interfaces orthogonal to the $z$-axis and filled with a nematic liquid crystal (LC) (see. Fig.1). An
externally applied electric field ${\bf E}_S$, lying in the $xz$ plane and forming an angle $\phi$ with the $z$-axis, is able to align the LC
molecules so that the LC homeotropic director is parallel to the unit vector $\hat{\bf n} = \sin \phi \hat{\bf e}_x + \cos \phi \hat{\bf e}_z$. As a
consequence the LC optically behaves as a uniaxially anisotropic crystal (with optical axis along the direction $\hat{\bf n}$) and its permittivity
tensor is $\epsilon^{(LC)} = R \: \textrm{diag} \left(\epsilon^{(LC)}_o,\epsilon^{(LC)}_o,\epsilon^{(LC)}_e\right) R^{-1}$ where $\epsilon^{(LC)}_o$
and $\epsilon^{(LC)}_e$ are the ordinary and extraordinary permittivities, respectively and
\begin{equation}
R = \left( \begin{array}{ccc} \cos \phi & 0 & \sin \phi \\ 0 & 1 & 0 \\ -\sin \phi & 0 & \cos \phi \end{array} \right)
\end{equation}
\begin{figure*}
\centering
\includegraphics*[width=0.8\textwidth]{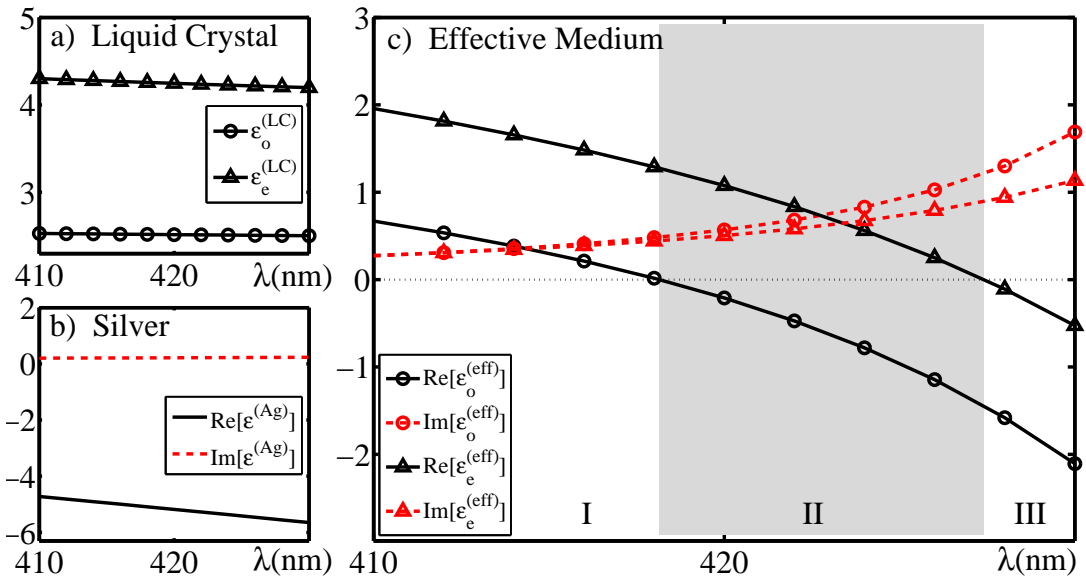}
\caption{Dielectric permittivities as a function of the vacuum wavelength $\lambda$. (a) Ordinary and extraordinary permittivities of the LC. (b)
Real and imaginary parts of the silver permittivity. (c) Real and imaginary parts of the ordinary and extraordinary permittivities of the homogeneous
effective medium. The shaded region corresponds to the spectral range where the effective permittivity tensor is indefinite.}
\end{figure*}
is the $y$-axis rotation matrix. The slab hosts silver nanoparticles of radius $r = 10$ nm which are dispersed within the LC with the volume filling
fraction $f$. If both the the radiation wavelength $\lambda$ and the slab thickness $L$ are much greater than the nanoparticles radius ($\lambda \gg
r$ and $L \gg r$), the overall optical response of the slab can be described by means of the effective medium approach suitably extended to encompass
anisotropic constituents \cite{Sihvo}. According to this approach, since silver nanoparticles are optically isotropic with scalar permittivity
$\epsilon^{(Ag)}$, the overall effective permittivity tensor is $\epsilon^{(eff)} = R \: \textrm{diag} \left(\epsilon^{(eff)}_{o} ,
\epsilon^{(eff)}_{o} , \epsilon^{(eff)}_{e} \right) R^{-1}$ so that the homogeneous effective medium is a uniaxial crystal with optical axis along
the direction $\hat{\bf n}$ and its ordinary ($j=o$) and extraordinary ($j=e$) effective permittivities are
\begin{equation} \label{effperm}
\epsilon^{(eff)}_{j} = \left[1 + f  \frac{ \left(\epsilon^{(Ag)}-\epsilon^{(LC)}_{j}\right)}{\epsilon^{(LC)}_{j}+ (1-f) N_{j} \left(
\epsilon^{(Ag)}-\epsilon^{(LC)}_{j} \right) }  \right] \epsilon^{(LC)}_{j}
\end{equation}
where the depolarization factors $N_{j}$ are
\begin{equation}
N_{j} = \int_0 ^{+\infty} ds \frac{\epsilon^{(LC)}_j}{ 2 \sqrt{\left(1+s\epsilon^{(LC)}_o\right)^2 \left(1+s\epsilon^{(LC)}_e\right)}}
\frac{1}{\left(1+s\epsilon^{(LC)}_j\right)}.
\end{equation}
The permittivities of Eq.(\ref{effperm}) are the anisotropic generalization of the standard Maxwell Garnett mixing rule (to which Eq.(\ref{effperm})
reduces in the isotropic limit where $N_j= 1/3$), the depolarization factors accounting for the effect of the anisotropy. Note that, as in the
isotropic situation, the nanoparticles radius does not appear in Eq.(\ref{effperm}) so that the mixture design have to be performed by tuning the
volume filling fraction $f$.

In order to discuss a feasible example of mixture, we have chosen SG-1 \cite{Gauza} as nematic LC due to its high birefringence and we have reported
its ordinary and extraordinary permittivities in Fig.2(a) as functions of the vacuum wavelength $\lambda$ in the spectral range between $410$ and
$430$ nm. For the silver permittivity we have used the Drude model $\epsilon^{(Ag)} = \epsilon_\infty - \omega_p^2 / ( \omega^2 + i \Gamma \omega )$
with $\epsilon_\infty = 4.56$, $\omega_p = 1.38 \cdot 10^{16} \: \rm{s^{-1}}$ and $\Gamma = 0.1 \cdot 10^{15} \: \rm{s^{-1}}$ \cite{Palik} and we
have plotted the real and imaginary parts of $\epsilon^{(Ag)}$ in Fig.2(b) for $\lambda$ between $410$ and $430$ nm. Note that we have used a
scattering rate $\Gamma$ which is an order of magnitude higher than that of bulk silver in order to accounting for the surface scattering (size
effect) occurring in the considered nanometric sized particles. After setting $f=0.05$ for the nanoparticles volume filling fraction, we have
evaluated from Eq.(\ref{effperm}) the principal permittivities of the effective medium and we have plotted their real and imaginary parts in
Fig.2(c). Note that for wavelengths between $418$ and $427.5$ nm, $\textrm{Re} \left( \epsilon^{(LC)}_e \right) > 0$ and $\textrm{Re} \left(
\epsilon^{(LC)}_o \right ) < 0$ so that the effective medium has indefinite permittivity. Combining this result with the fact that the direction of
the optic axis can be varied through the externally applied electric field, we conclude that the proposed LC/silver nanoparticles mixture behaves,
within the limits of validity of the effective medium theory, as a tunable hyperbolic medium.

\begin{figure*}
\centering
\includegraphics*[width=0.9\textwidth]{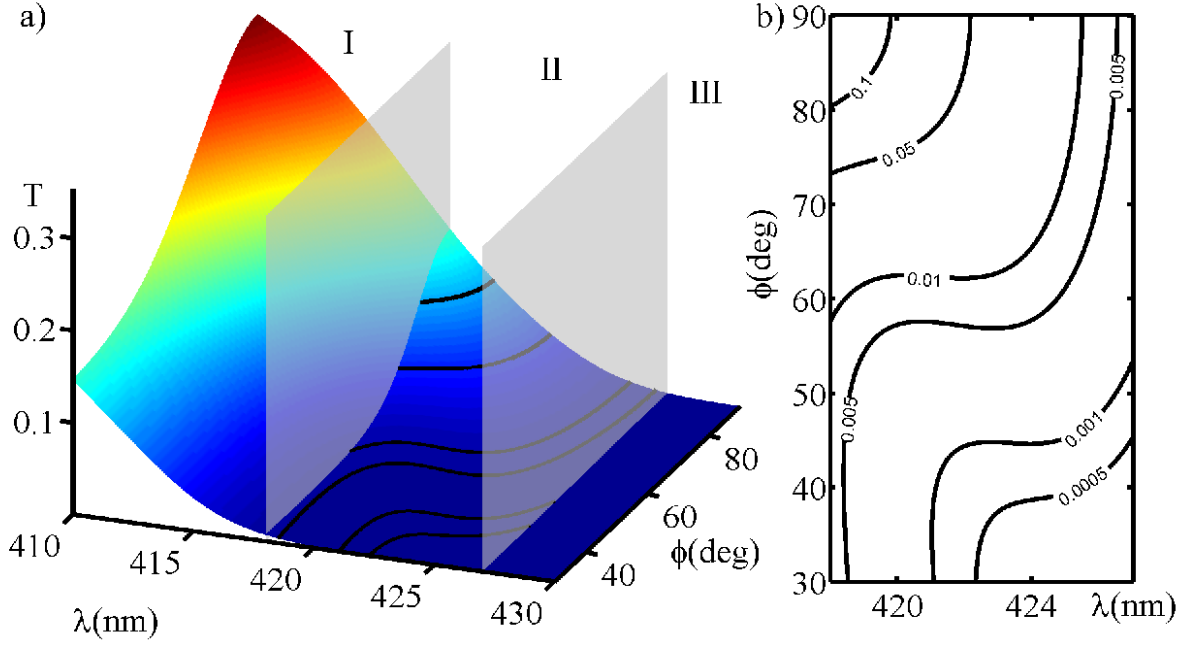}
\caption{(a) Plot of the transmissivity of the device reported in Fig.1 (evaluated using the effective medium model for the LC/nanoparticels mixture)
as a function of the vacuum wavelength of the incident radiation and the angle $\phi$. (b) Level plot of the transmissivity of panel (a) restricted
to region II where the effective medium has indefinite permittivity.}
\end{figure*}

\section{Optical switch}
The intriguing optical properties of hyperbolic media combined with the tunability, allow the above described LC/nanoparticles mixture to be used for
conceiving optical devices. As an example, consider the setup of Fig.1 where the LC slab with dispersed nanoparticles has thickness $L=300$ nm and it
is sandwiched between two $d=70$ nm thick silica layers (of permittivity $\epsilon^{(S)} = 2.088$). If a monochromatic plane wave is made to impinge
orthogonally, from vacuum, onto the device interface (as sketched in Fig.1), it is evident that we are considering an active device since its optical
transmissivity can be altered by varying the direction of the externally applied electric field ${\bf E}_S$. Using the effective medium approach
discussed in Sec.2 to model LC/nanoparticles mixture, the device transmissivity is easily evaluated and it is reported in Fig.3(a), as a function of
both the vacuum wavelength $\lambda$ and the angle $\phi$ defining the direction of the applied electric field. For clarity purposes, in Fig.3(a) we
have reported two transparent gray planes corresponding to the two wavelengths $418$ and $427.5$ nm between which the effective medium predicts that
the mixture is hyperbolic. The most striking feature of Fig.3(a) is the existence of three relevant spectral regions, I) $\lambda < 418$ nm, II)
$418$ nm $<\lambda<427.5$ nm and III) $\lambda > 427.5$ nm, where the transmissivity is characterized by qualitatively different behaviors.
Specifically within regions I and III, the slab is almost transparent and completely opaque, respectively whereas, within region II the
transmissivity strongly depends on $\phi$ to the point that $T \simeq 0$ for $\phi \simeq 0$ and, for increasing $\phi$, it gradually increases
eventually reaching its maximum (dependent on $\lambda$) for $\phi = \pi/2$ (see Fig.3(b) where the level plot of the transmissivity of region II is
plotted).
\begin{figure*}
\centering
\includegraphics*[width=0.9\textwidth]{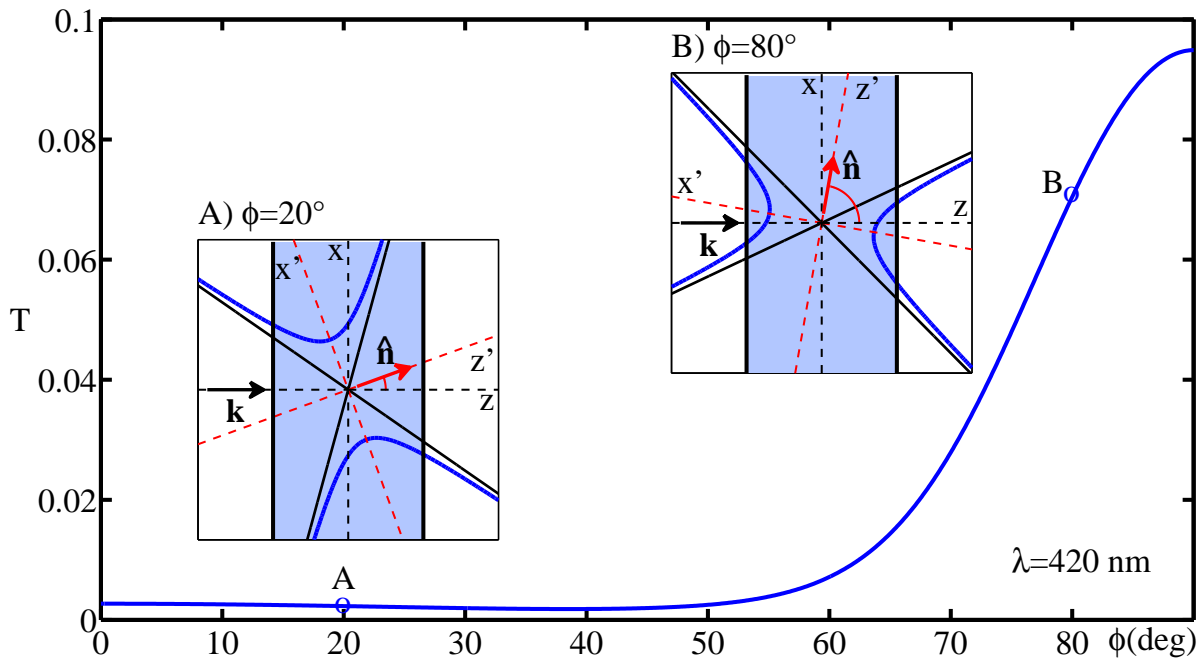}
\caption{Transmissivity $T(\phi)$ for $\lambda = 420$ nm extracted from Fig.3(a). In the insets, the extraordinary waves hyperbolas of
Eq.(\ref{disp}) are plotted in the two relevant situations (A) $\phi = 20^\circ$ and (B) $\phi = 80^\circ$ (regarding the permittivities as real
quantities).}
\end{figure*}
Therefore, within region II, the considered structure can be regarded as a very thin optical switch which, being driven by the external electric
field, is able to completely stop or allow the transmission of the light flow. Physically, such a behavior is readily grasped by noting that within
region I and III (see Fig.2(c)) the effective medium behaves as a uniaxial crystal and an anisotropic conductor, respectively and therefore, within
the same two regions, the plane wave impinging from vacuum excites homogeneous and evanescent waves, respectively. On the other hand, within region
II the effective medium is indefinite so that, depending on the value of $\phi$, the normally impinging plane wave incoming from vacuum is able to
excite either homogeneous or evanescent waves. In order to clarify this crucial point, we start from the dispersion relation of extraordinary plane
waves
\begin{equation} \label{disp}
\frac{k_x'^2}{\epsilon^{(eff)}_{e}}+\frac{k_z'^2}{\epsilon^{(eff)}_{o}} = k_0^2
\end{equation}
where $k_x'$ and $k_z'$ are the components of the wave vector along the principal axis of the uniaxial crystal and $k_0 = 2\pi / \lambda$. Since the
plane wave from vacuum normally impinges onto the device interface (see Fig.1), conservation of momentum requires that the wave vector of waves
excited within the anisotropic slab is ${\bf K} = K \hat{\bf e}_z$, i.e. it is along the $z$-axis. Therefore the wavevector components along the
principal axis are $k_x'=-K \sin \phi$ and $k_z'=K \cos \phi$ which, inserted into Eq.(\ref{disp}) yield
\begin{equation} \label{kappa}
K(\phi)=k_0 \sqrt{\frac{\epsilon^{(eff)}_{o} \epsilon^{(eff)}_{e}}{\epsilon^{(eff)}_{o} \sin^2 \phi + \epsilon^{(eff)}_{e} \cos^2 \phi}}
\end{equation}
from which the impact of the angle $\phi$ on the excited waves is particularly evident. Even though we are not neglecting absorption (see the
imaginary parts of the permittivity in Fig.2(c)), it is convenient here to regard the principal permittivities as real quantities with
$\epsilon^{(eff)}_{o}<0$ and $\epsilon^{(eff)}_{e}>0$. Therefore, from Eq.(\ref{kappa}), we readily obtain that $K(0)$ is imaginary (evanescent
waves) whereas $K(\pi /2)$ is real (homogeneous waves), the switch between the two kinds of waves occurring at the angle $\phi = \arctan
\sqrt{-\epsilon^{(eff)}_e/\epsilon^{(eff)}_0}$. This explains the dependence of $T$ on $\phi$ within region II of Fig.3, situation where absorption
has not been neglected, its only effect being the smoothing of the dependence of $T$ on $\phi$.
\begin{figure*}
\centering
\includegraphics*[width=0.9\textwidth]{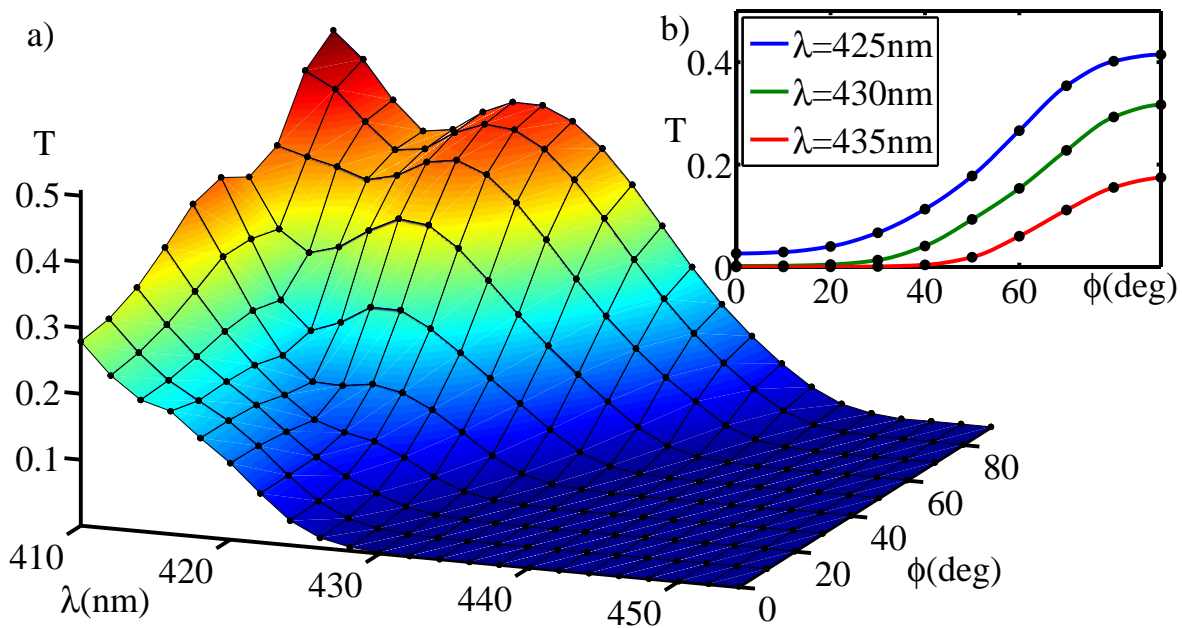}
\caption{(a) Device transmissivity $T$ as a function of the vacuum wavelength $\lambda$ and the angle $\phi$ evaluated through full wave simulations.
(b) Transmissivity $T(\phi)$ extracted from panel (a) at three different wavelengths where the mixture displays indefinite character.}
\end{figure*}

The just described mechanism is pictorially sketched in Fig.4 where we have reported the transmissivity $T(\phi)$ for $\lambda = 420$ nm extracted
from Fig.3(a). In the insets $A)$ and $B)$ of Fig.4 we have plotted the hyperbolas of Eq.(\ref{disp}) in the two relevant situations (A) $\phi =
20^\circ$ and (B) $\phi = 80^\circ$ (regarding the permittivities as real quantities). The direction of the applied electric field is able to
physically rotate the medium hyperbola and, as a consequence, in the two situations $\phi= 20^\circ$ and $\phi= 80^\circ$ the hyperbola does not
intersect and intersects the $z$-axis, respectively, thus correspondingly allowing the normally impinging vacuum wave to excite evanescent and
homogeneous waves. The switch between the two kinds of waves occurs when the hyperbola asymptote is nearly normal to the slab interface and this
occurs for $\phi$ close to $\arctan \sqrt{-\epsilon^{(eff)}_e/\epsilon^{(eff)}_0}$.

\section{Full wave simulations}
In order to check the homogenization of the LC/nanoparticles mixture and the device switching functionality, we have performed 3D full-wave
simulations by means of a finite-element solver, using the numerical values of the parameters (constituent permittivities, nanoparticles radius and
volume filling fraction, layers thicknesses, etc.) considered so far. In Fig.5(a) we have plotted the resulting transmissivity $T$ as a function of
the vacuum wavelength $\lambda$ and the angle $\phi$. It is worth noting that, even though the surface of Fig.5(a) does not strictly coincide with
that of Fig.3(a), it is however evident that the actual structure is almost transparent and practically opaque for wavelengths $\lambda$ smaller than
$430$ nm and greater than $445$ nm, respectively. In addition, in the spectral range between $430$ and $445$ nm the transmissivity strongly depends
on the angle $\phi$ and, at a given $\lambda$, $T(\phi)$ is a monotonically increasing function of $\phi$ (see Fig.5(b)). As a consequence, even
though the spectral ranges are slightly different from those predicted by the effective medium approach, the sample behavior changes, for increasing
radiation wavelengths, from that of a dielectric uniaxial crystal to that of an anisotropic conductor with an intermediate spectral region where the
medium is hyperbolic. In Fig.5(b) we have plotted the transmissivity of the strucutre as a function of $\phi$ for three relevant wavelengths and the
switching functionality is particularly evident. Moreover, since the transmissivity spans the range between $0.02$ and $0.4$ for $\lambda=425$ nm, we
conclude that the performance of the actual switching device is better than that predicted by the effective medium theory (whose transmissivity range
is between $0$ and $0.15$).

The discrepancies between the predictions of the homogenized effective medium theory and those of the full-wave simulations are a consequence of the
fact that, for the chosen slab thickness $L=300$ nm, the edge effect due to the nanoparticles close to the slab interface can not be strictly
neglected. The derivation of the Maxwell Garnett mixing rule (whose anisotropic extension has been used in this paper) is based on the evaluation of
the polarizability of an isolated nanoparticle experiencing an external field which is uniform within its volume. For particle belonging to the
mixture bulk, these requirements are fulfilled if the nanoparticle size is much smaller than the wavelength and if the nanoparticles volume filling
fraction is very small. However, for nanoparticles close to the sample interface the above polarizability evaluation is wrong since it does not take
into account the additional field existing within the nanoparticle which is due to surface polarization charges (i.e. to the discontinuity of the
electromagnetic material properties at the interface) \cite{Stepa}. Since the corrections to the permittivity scales as $(r/z)^3$ ($z$ being the
distance from the interface) the nanoparticles responsible for the edge effect are those whose distance from the interface is smaller than $40$ or
$50$ nm (four or five times its radius $r=10$ nm), so that we conclude that roughly a fourth of the considered slab volume can not be described
through the Maxwell Garnett approach.

\section{Conclusions}
We have proposed a novel way for synthesizing indefinite media by dispersing metal nanoparticle within a highly birefringent liquid crystal. In
addition to its simplicity and feasibility, the proposed mixture is also tunable since an externally applied electric field is able to change the
overall effective permittivity tensor. Tunability of an indefinite medium is a very important property since, as we have shown, the ability of
altering the orientation of the extraordinary wave hyperbola allows one to literally decide whether the electromagnetic field within the medium (as
due to a give external excitation) has to be composed of propagating homogeneous or an evanescent waves. We have exploited such a property
characterizing tunable indefinite media to design a nanometric thick optical switch and investigated its functionality both through the effective
medium approach and through full wave simulations. The effective medium approach has the sufficient simplicity to highlight the basic physical
mechanisms supporting the device functionality. On the other hand, full wave simulations yield results which are slightly different from those
predicted by the effective medium approach and the discrepancy are related to the very small device thickness we decided to consider. However, from
the point of view of the device functionality, the results of the full wave simulations are even better than those of the effective medium approach.

\end{document}